\documentclass[prl,aps,twocolumn,floatfix]{revtex4}

\usepackage{amssymb,graphicx}
\usepackage{epsfig}
\usepackage[usenames]{color}

\newcommand{\secintro}{I}
\newcommand{\secapproach}{II}
\newcommand{\secphysics}{III}
\newcommand{\secresults}{IV}
\newcommand{\secconclusion}{V}

\begin{document}

\title{Magnetized Neutron Star Mergers and Gravitational Wave Signals}

\author{Matthew Anderson${}^1$, Eric W. Hirschmann${}^2$,
Luis Lehner${}^3$, Steven L. Liebling${}^4$, Patrick M. Motl${}^3$,
David Neilsen${}^2$, Carlos Palenzuela${}^{3,5}$, Joel E. Tohline${}^3$}

\affiliation{
${}^1$Department of Mathematics, Brigham Young
University, Provo, UT 84602,\\
${}^2$Department of Physics and Astronomy, Brigham Young
University, Provo, UT 84602,\\
${}^3$Department of Physics and Astronomy, Louisiana State
University, Baton Rouge, LA 70803-4001, \\
${}^4$Department of Physics, Long Island University--C.W. Post Campus,
Brookville, NY 11548\\
${}^5$Max-Planck-Institut f\" ur Gravitationsphysik,
Albert-Einstein-Institut, Golm, Germany
}

%
%
\begin{abstract}
We investigate the influence of magnetic fields upon the dynamics of, and 
resulting gravitational waves from, a binary neutron 
star merger in full general relativity coupled to 
ideal magnetohydrodynamics~(MHD).  
We consider two merger
scenarios, one where the stars have aligned poloidal magnetic fields 
magnetic fields and one without.  Both mergers result
in a strongly differentially rotating object.
In comparison to the non-magnetized scenario, the aligned magnetic fields
delay the full merger of the stars.   During and after
merger we observe phenomena driven by the magnetic field, including 
Kelvin-Helmholtz 
instabilities in shear layers, winding of the field lines, and transition 
from poloidal to toroidal magnetic fields. These effects
not only mediate the production of electromagnetic radiation, but also can  
have a strong influence on the gravitational waves. Thus, there are
promising prospects for studying such systems with both types of waves.
\end{abstract}

\maketitle

%
%
\noindent{\bf{\em \secintro. Introduction:}} 
To the long list of spectacular electromagnetic signals which researchers have
employed to study our universe, a new counterpart of gravitational
origin will soon be added. Gravitational waves will both complement our ability
to probe a variety of sources and in many cases provide the only option.
These waves
will likely be detected in the near future by a new generation of
laser interferometers~\cite{Frey:2007zz,Willke:2007zz,Acernese:2005yh}.
Among
their most promising
sources
are merging binaries composed of black holes
or neutron stars.  These waves carry information about the
dynamic, strong gravitational field near the binary.  When one member of
the binary is a neutron star, these waves can have a richer structure and
carry additional information about the matter.  For example, analysis of
such mergers can potentially yield constraints on the nuclear equation of state and, as we
show here, key insights on the stars' magnetic fields.
The combination of gravitational wave data with other astronomical observations
in the electromagnetic spectrum
will provide an enhanced understanding of many energetic, strongly gravitating
phenomena, such as (short) gamma ray bursts.

Precise theoretical models are required to extract information from gravitational waves.
Recently, significant progress has been made in modeling the radiation
expected from binary black holes, binary neutron stars, and black hole-neutron
star binaries.  However, much less is known about how magnetic
fields may alter the dynamics of such mergers, whose effects would 
be encoded in the gravitational waveforms.  
Magnetic fields are present in many astrophysical systems, and are
important in several interesting phenomena, such as AGNs, pulsars,
gamma ray bursts, etc.
Observational evidence indicates that neutron stars have some of
the strongest known magnetic fields, with estimated field strengths ranging 
from $10^{12}$~G to as large as $10^{16}$~G in 
magnetars~\cite{Duncan:1992hi}. Magnetic field effects in merging/collapsing
scenarios have been investigated in related systems. For instance,  
perturbation theory has been employed to examine magnetized dust 
collapse~\cite{Sotani:2007jv},
Newtonian gravity to study magnetized binary neutron star 
mergers~\cite{pricerosswog}, Post-Newtonian (PN) calculations to inspect
the pre-merger dynamics~\cite{2000ApJ...537..327I}
and full general relativity (reduced to 2D by symmetries) 
to analyze axisymmetric differentially rotating star 
collapse~\cite{Duez:2005cj}.
We study the merger in full, 3D general relativity with a framework 
able to study the fluid's behavior and
extract the gravitational waves in the wave-zone. 
Clearly this system is
highly relevant in the mechanism underlying short gamma ray burst phenomena.
However, since we do not include radiation transport 
our focus in the present work is on the gravitational
wave signature and the fluid's behavior in a scenario probing 
the most extreme possible influence of the magnetic field.

Magnetic fields can have a strong influence on the dynamics of  stars 
by introducing a number of instabilities (see e.g.~\cite{Spruit:2007bt}). 
These instabilities allow for energy exchange 
between the fluid and magnetic modes
and can redistribute the fluid's angular momentum through different 
mechanisms (e.g., magnetic winding and braking, magneto-rotational 
instabilities), depending on the strength of the magnetic field and the 
typical dynamical time scales~\cite{Spruit:1999,Piro:2007}.  
These effects can have a 
non-trivial effect on the (time dependent) multipolar structure of the source, 
and consequently on the resulting gravitational waves. 
A detailed understanding of magnetic effects in binary mergers and stellar
collapse will provide valuable information not only for deciphering the
observed gravitational waves,
but they might also influence the design and tuning of future advanced 
detectors~\cite{advligo,Mandel:2007hi}.

In the present work we investigate possible magnetic effects in general
relativity by simulating a merger of two compact stars with
and without the presence of a magnetic field. 
We study the merger in full, 3D general relativity with a 
framework able to study the fluid's behavior and
extract the gravitational waves in the wave-zone. 


%
%

\noindent{\bf{\em \secapproach. Overview of the numerical approach:}} 
We model the neutron star material using relativistic ideal magnetohydrodynamics (MHD),
coupled to the full Einstein equations of general relativity to 
represent accurately the strong gravitational effects during the merger.
Our numerical techniques for solving these coupled equations have been
tested previously~\cite{Anderson:2006ay,binaryNS,Palenzuela:2006wp,Liebling}.
Some essential
elements of our implementation include the following. 
We use a computational 
infrastructure that provides distributed 
Adaptive Mesh Refinement (AMR)~\cite{had_webpage,Liebling,Lehner:2005vc}. 
with full sub-cycling in time, with a novel
treatment of artificial boundaries~\cite{Lehner:2005vc}.  
The refinement regions are determined 
by the truncation error of the solution during the evolution.  
We apply refined boundary conditions 
at the outer boundary to ensure consistency of the solution over
the computational domain~\cite{Palenzuela:2006wp,RLS07} and adopt
high resolution shock-capturing methods together with divergence cleaning
to control the no-monopole constraint~\cite{Anderson:2006ay}.
Our methods to solve the Einstein equations~\cite{Palenzuela:2006wp,RLS07} 
and MHD equations~\cite{Anderson:2006ay} are described elsewhere.

%
%
\noindent{\bf{\em \secphysics. Physical set-up:}} 
To begin exploring possible magnetic effects in binary systems, 
we compare two different binary neutron star
mergers, one with a magnetic field and one without.
All other initial data are identical.  
For the magnetized stars, we choose a 
particular case in which each star initially has a
poloidal magnetic field anti-aligned with the orbital angular momentum.
Because the stars' magnetic fields are themselves aligned,
this configuration induces a magnetic repulsion as the stars 
approach.  Strong field repulsion should 
impact the dynamics of the system and our results are compared with PN
results presented in~\cite{2000ApJ...537..327I}.

We construct the initial data with two identical
neutron stars with zero spin angular momentum.  
Each star has a mass of
$M=0.89~M_{\odot}$, a radius of $16.26$~km, and a central density of
$3.24 \times 10^{14} \; {\rm g}/{\rm cm}^3$.
The stars are initially separated by $60$~km and are given initial tangential
velocities slightly below the Keplerian value.  We adopt a simple
polytropic equation of state 
choosing $\Gamma=2$ (polytropic index $n=1$) to model stiff nuclear  matter~\cite{binaryNS}. 

The poloidal magnetic field is calculated from the vector 
potential
$   A_\varphi = \varpi^2 \max(P-P_{\rm vac}, 0), $
where $P$ is the pressure, $\varpi$ the cylindrical radius, and
$P_{\rm vac}/c^2 \simeq 10^7$~gm/cm$^3$~\cite{Shibata:2006hr}.   
The field is 
initially confined to the stellar interior, and its maximum magnitude 
is approximately $9.6 \times 10^{15}$~G.
While this value is large for astrophysical stars (excepting magnetars),
it is small compared to the gravitational binding energy of the stars.  
As the individual stars do not rotate differentially, the initial field
remains essentially constant until the stars are close enough for strong
interactions.

The initial data are evolved in a cubical computational domain given by 
$[-1540 \,\rm{km},1540 \,\rm{km}]$
in each direction with an  AMR configuration having
$7$ levels of refinement for the finest resolved simulation 
($5$ and $6$ levels of refinement were used for comparison).
The highest resolution region follows
each star and has a grid spacing of $\Delta = 0.46$~km.
We extract the gravitational wave information and compute the strain $\{ h_{+}, h_{\times} \}$.
Observers are placed at three different radii from the center
of mass, namely $r=\{440,590,740\}$~km. These correspond to 
$r\simeq\{5,7.5,10\} \lambda_{\rm max}$, where $\lambda_{\rm max}$ is 
the largest wavelength in the observed gravitational wave. The extraction takes
place in the wave-zone, which is evident by the excellent 
agreement among the computed waveforms.

\noindent{\bf{\em \secresults. Results:}} 
The evolution of both binary pairs progresses quite similarly.
The stars' orbits shrink as energy is lost
through gravitational wave emission. As they merge,
a bar is formed, which itself is unstable.  The bar expels matter and collapses to 
form a differentially rotating, hypermassive star. While rotation
initially supports the star, the continued loss of energy through wave
emission causes it eventually to collapse, forming a black hole.
Comparing these two mergers, however, we see that the magnetic field
strongly influences the details and time scales of the collapse, resulting 
in a solution that is clearly distinguishable from the unmagnetized 
case.

\begin{figure}[h]
 \begin{center}
 \epsfig{file=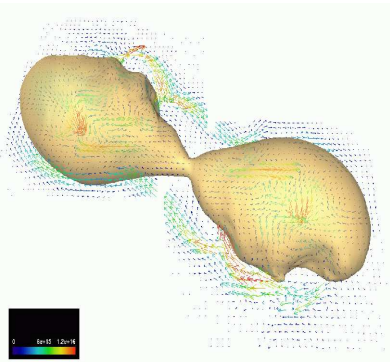,height=4.8cm,width=5.2cm}
 \caption{Fluid density isopycnic (at $\rho=6 \times 10^{13} g/cm^3$) and 
 magnetic field distribution (in a plane slightly above
 the equator) at $t=4.4ms$. 
 Kevin-Helmholtz instability in the shear layer has produced
 circulation cells that amplify the toroidal magnetic field.
 The cavities at both trailing edges are due to magnetic pressure
 inducing buoyancy. }
 \label{fig:field}
 \end{center}
 \end{figure}

\begin{figure}[ht]
 \begin{center}
 \epsfig{file=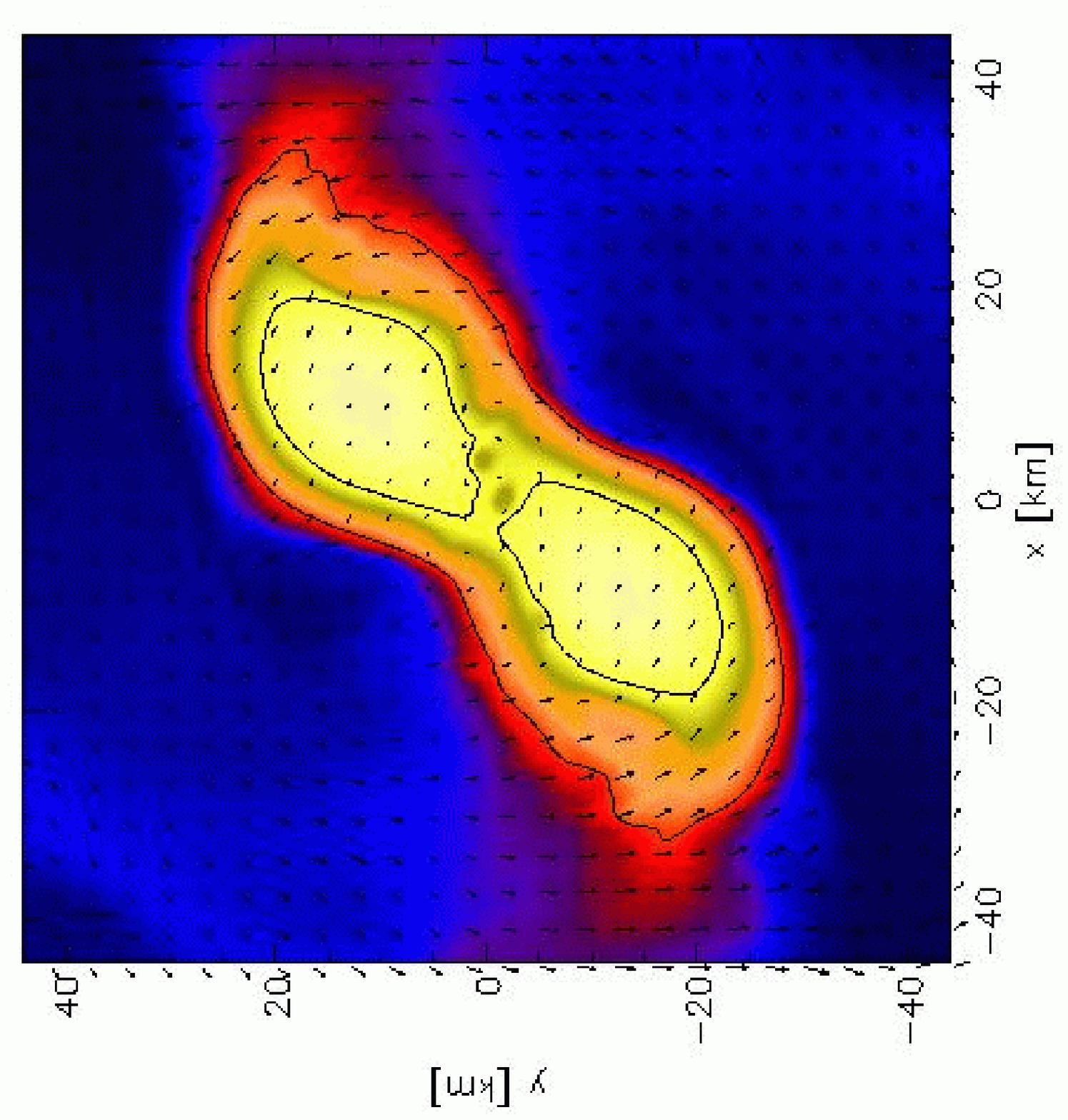,height=4.2cm}
 \epsfig{file=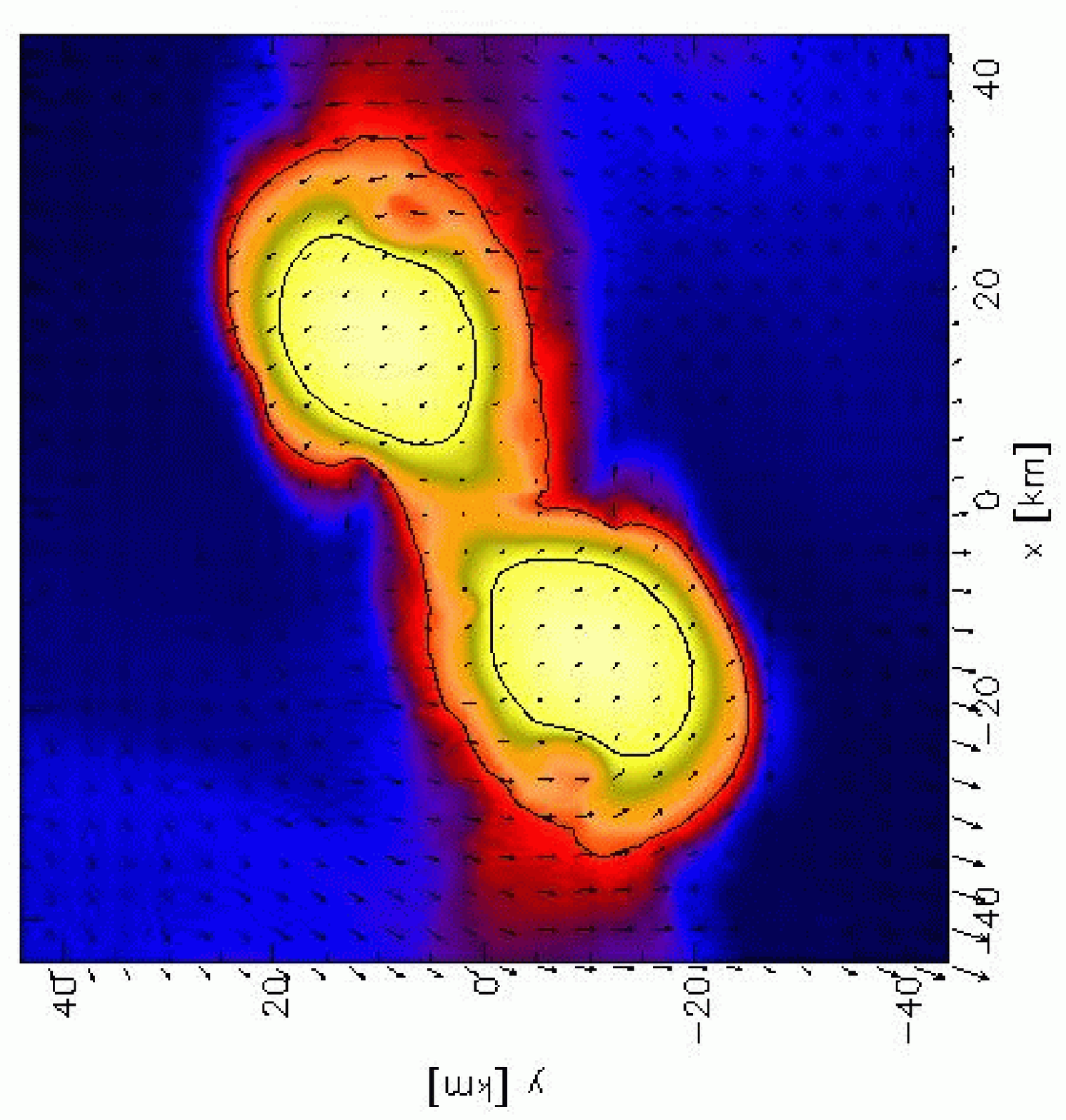,height=4.2cm}
 \epsfig{file=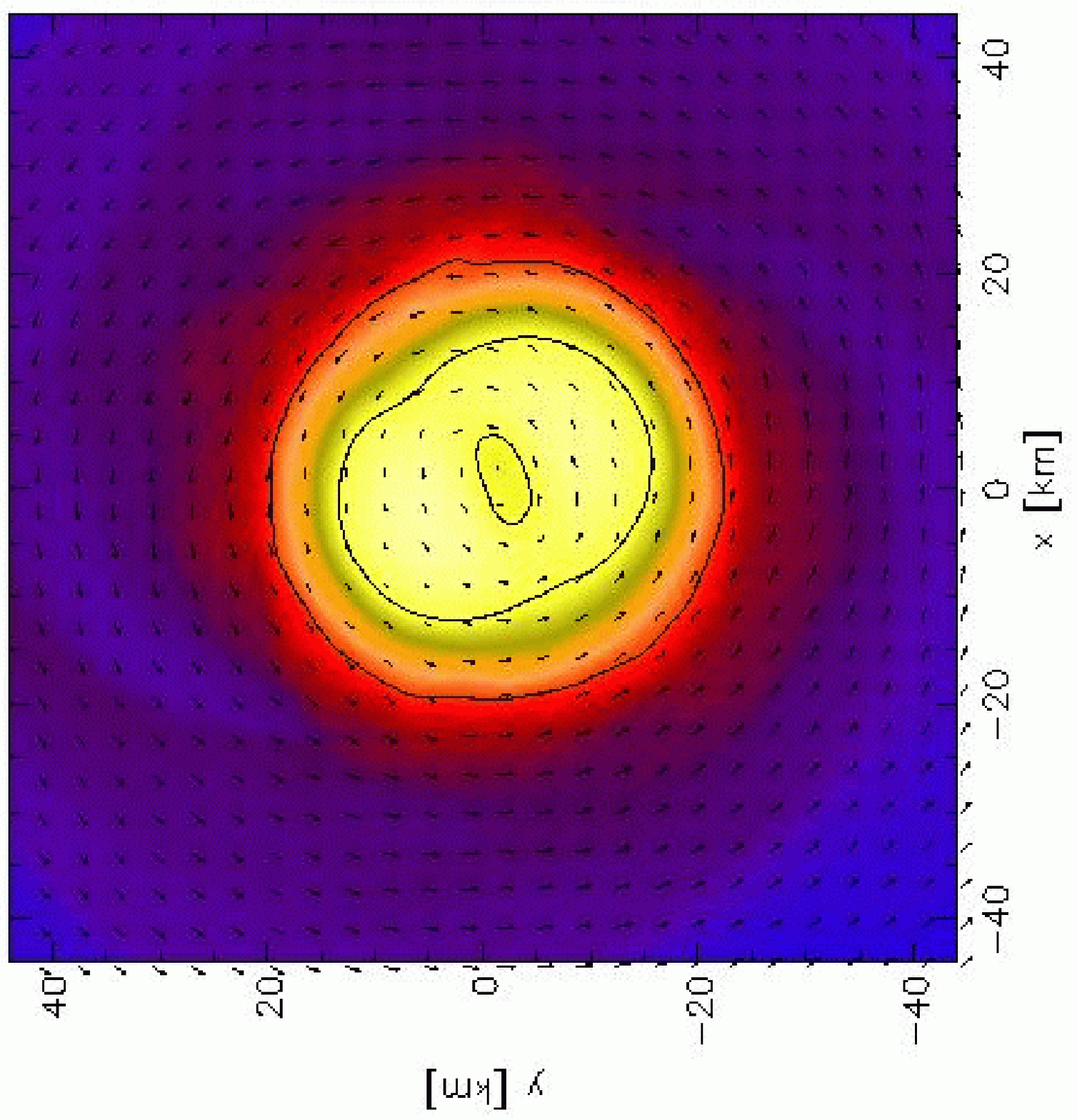,height=4.2cm}
 \epsfig{file=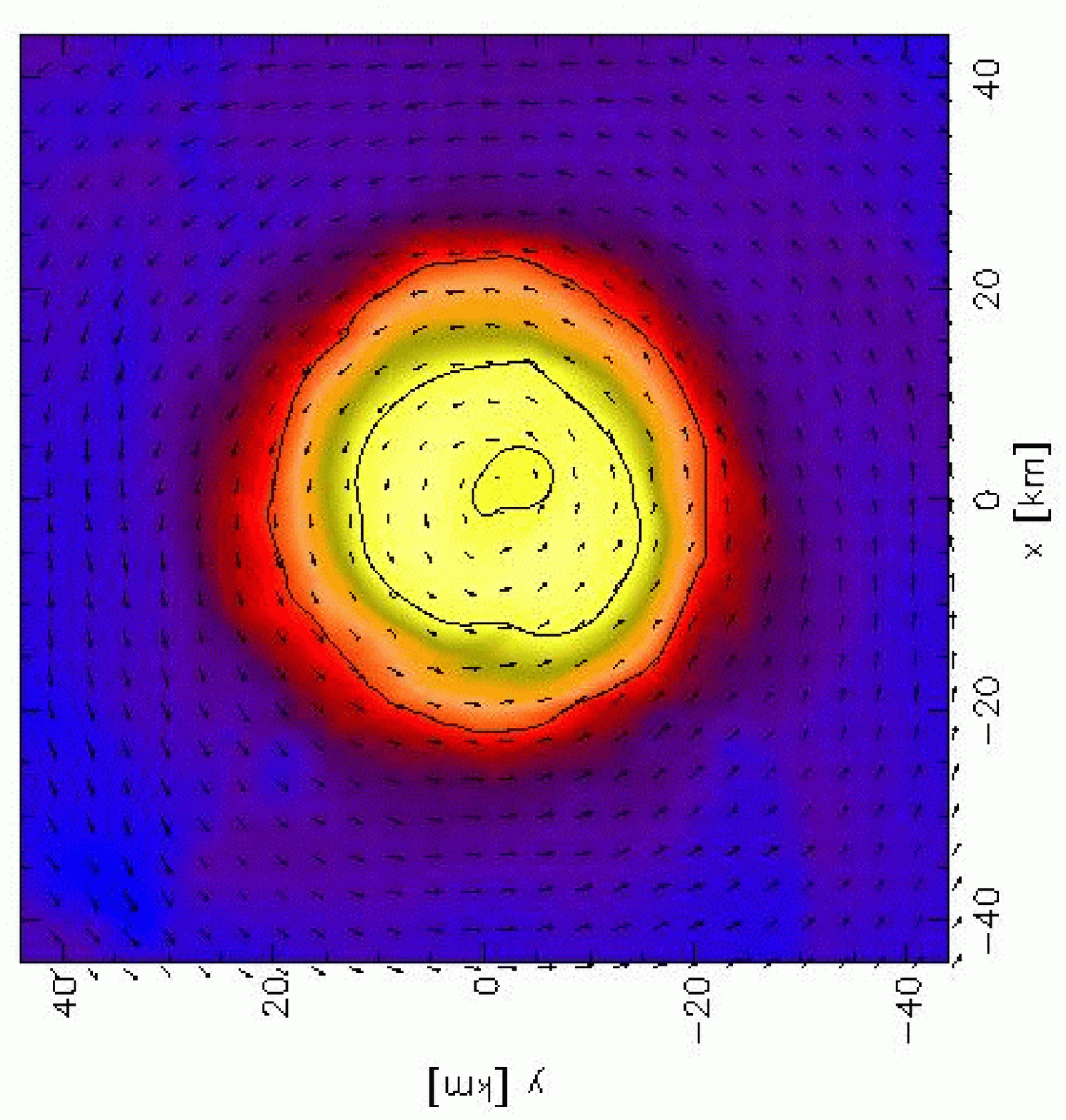,height=4.2cm}
 \caption{Equatorial density contours and velocity
 field at two times for both 
 the non-magnetized (left column) and magnetized cases (right column). 
 The top frames are at $t=4.9$~ms,
 clearly shows the delayed merger caused by  magnetic field repulsion. 
 The bottom frames are at $t=18.6$~ms. With no magnetic field,
 the final differentially rotating star displays a more bar-like 
 structure (m=2), while the magnetized star is considerably more spherical 
 with a slight off-set.} 
 \label{fig:fluidsnapshots}
 \end{center}
 \end{figure}

\begin{figure}[h]
 \begin{center}
 \includegraphics[angle=-90,width=0.4\textwidth]{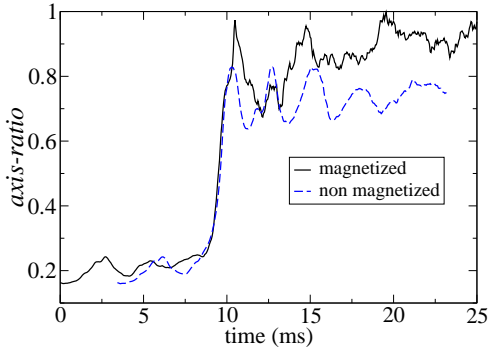}
 \caption{
 Equatorial axis-ratio (semi-minor/semi-major) vs. time for the non-magnetized (blue-dashed line) and
 magnetized (black-solid line) measured at a density of  
$1.24 \times 10^{14}$~g/cm$^3$.
 the results corresponding to the non-magnetized case has been shifted in
 time by $t=3.5$~ms so both mergers coincide. 
 Notice how the magnetized case 
 approaches much quicker the unit value while the non-magnetized 
 remains about $\simeq 0.78$. This translates into stronger amplitude in the produced
 gravitational waves.} 
 \label{fig:ratio}
 \end{center}
 \end{figure}

Initially, the two solutions are nearly identical.  
When the stars begin to disrupt and shed matter, initiating the merger phase,
the aligned magnetic fields of the stars repel each other, delaying
the final merger by 1-2 rotation periods with respect to the unmagnetized case. This delay is in agreement with 
post-Newtonian results presented in~\cite{2000ApJ...537..327I}. 
As the merger proceeds, a number of magnetically induced phenomena are 
observed~\cite{Spruit:1999,Spruit:2007bt,Piro:2007}.
First, a shear interface forms giving rise to Kelvin-Helmholtz instabilities which are
responsible for a rapid growth in the magnetic field strength, as shear 
energy is converted 
into magnetic energy (see Fig.~\ref{fig:field}).
Second, differential rotation of the resulting merged star winds
the magnetic field up, producing a toroidal field and linearly
amplifying it as the star rotates.
This trend saturates on the order of 
an Alfv\'en time scale, which in our case is $t_A \simeq 20 P_c$, where $P_c$ is the rotational 
period of the star's core.  In our simulation this period is about 
$P_c=0.735$~ms (cf.~\cite{Duez:2005cj}). 
Third, at least two types of Tayler instabilities are 
displayed~\cite{1973MNRAS.162...17T,Spruit:1999}. 
One is responsible for inducing an $e^{im\phi}$, 
$m=1$ perturbation~\cite{Spruit:1999}. 
The second one, induces a magnetic buoyancy
instability which sheds some portions of the star in low density regions 
(about $1\%$ of the star's maximum density) which manifest in cavities that can be observed
in the magnetic case but are absent otherwise.
 Finally, as the star collapses to a 
black hole,  the magnetic field strength grows due to the compression 
of the field lines. 
The effects mentioned above strongly impact the fluid's distribution.
Contrasting the magnetized case with the unmagnetized,
Figure~\ref{fig:fluidsnapshots} shows density contours while Figure~\ref{fig:ratio} displays
the ratio of the equatorial semi-minor and semi-major axes. 

\begin{figure}[h]
 \begin{center}
 \includegraphics[angle=-90,totalheight=0.2\textheight,width=0.42\textwidth]{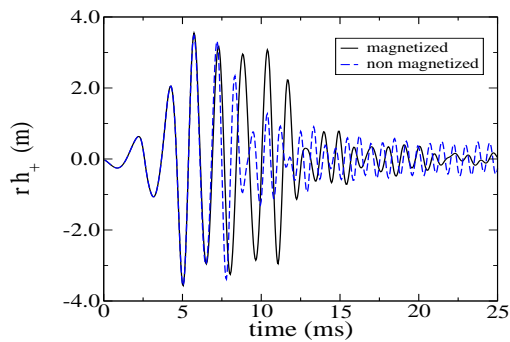}
 \caption{Gravitational wave signal $h_+$ ($l=2,m=2$ mode)
 for the magnetized (solid line) and unmagnetized (dashed) mergers.
 Before $t\simeq 7.5$~ms, both signals are essentially the same.
 After $t\simeq 7$~ms, however, the delayed merger of the magnetized stars
 results in a much stronger signal lasting until $t\simeq 12$~ms.
 After merger, the redistribution of angular momentum by the magnetic field
 results in a more axisymmetric remnant whose core
 rotates more slowly. As a result, at late times the non-magnetized merger 
 produces radiation with larger/higher amplitude/frequency 
 than its magnetized counterpart.
 \label{waveforms}
 }
 \end{center}
 \end{figure}

The combined effects of the magnetic field on the merger solution
have a strong influence on the gravitational waveforms. 
Figure~\ref{waveforms} shows the $l=2,m=2$ mode 
of $h_{+}$, the gravitational wave signal.
At early times before merger, the stars'
magnetic fields do not strongly interact, and the waveforms are
nearly identical.  The delayed merger is clearly visible in the large 
amplitude of $h_+$ which extends to $t\simeq 12$~ms, or 1-2 more cycles
than the non-magnetized case.  Magnetic effects in
the post-merger waveform are also evident: $h_+$ has both a smaller
amplitude and lower frequency, especially for $t\gtrsim 20$~ms.
The magnetic field transports angular momentum away from the center,
and as a result the rotation profile of the magnetized
case displays a lower/higher frequency at the inner/outer regions of 
the star and a rather uniform rotation profile in between together with
a more axisymmetric configuration
(see Figs.~\ref{fig:fluidsnapshots} and~\ref{fig:ratio}). Thus the core
is more axisymmetric and has a longer rotation period in the magnetized case
than in the non-magnetized scenario. It is important to note that while
our initial configuration has a large magnetic field, magnetic fields have
been observed to saturate near this value in merger simulations that begin with
weaker fields ($\simeq 10^{13}G$)~\cite{pricerosswog}.
The similarity in these post-merger field strengths and configurations leads
us to believe that the dynamics will demonstrate analogous behavior.
Finally, the frequency range of the obtained waveform
is about $2.7$--$6.4$~kHz. LIGO is sensitive to strains 
of $\simeq 4 \times 10^{-22}/{\sqrt{\rm Hz}}$ in this range~\cite{Frey:2007zz}.
The waves shown in Figure~\ref{waveforms} have amplitudes
comparable to this noise at distances of $2$ ($4$) Mpc
for the non-magnetic (magnetic) cases.
Advanced LIGO would be able to detect these
sources ten times farther, including the Virgo cluster. 


%
%
\noindent{\bf{\em \secconclusion. Conclusion:}} 
We have presented
a simulation of magnetized neutron stars with full 3D GR and MHD 
with sufficient
resolution to capture key magnetically-driven phenomena. 
We observe the transfer of
shear energy into magnetic energy and magnetic winding, which are 
responsible for
the creation and amplification of a toroidal field in the resulting
hypermassive star.
The strong differential rotation of this star
transforms the poloidal field into a toroidal field, the former decreasing by
an order of magnitude. The toroidal field redistributes angular
momentum, resulting in an inner core rotating more slowly than the 
non-magnetized case. While some of these effects have been seen 
to varying degrees in recent 
models (2D-axisymmetric simulations in GR~\cite{Shibata:2006hr},
3D SPH simulations without GR \cite{pricerosswog}), our implementation 
allows us to observe the magnetic interactions
during the merger, consistently accounting for the tightening of the orbit due
to the emission of gravitational waves, and does not restrict magnetic effects due
to symmetries. The merger delay reported here
has a strong influence in the resulting waveforms, since 
they have more relatively high-amplitude cycles. 
Additionally, the subsequent waveform displays observable differences 
in amplitude and frequency due to different magnetic instabilities. 
In our results, the inspiral (chirp) phase displays no significant 
difference but the merger and after-merger epochs do.
In fact a cross-correlation between the two waveforms is equal to $0.98$, 
$0.63$, and $0.59$,
when integrating from $t=0$ to $t=7.5$, $t=12.5$, and $t=25$~ms, respectively.
Searches for these types of magnetic effects in gravitational waves can give (upper-limit)
population estimates  of these and related binaries. Much more work will be required
to map out the possible configuration space of field strengths and 
orientations, but the
tantalizing differences observed indicate interesting results can 
be obtained and linked to gravitational and electromagnetic observations.

%
%
\noindent{\bf{\em Acknowledgments:}}
We would like to thank L.~Bildsten, O.~Blaes, J.~Frank, 
G.~Gonzalez, J.~Pullin, I.~Olabarrieta and O.~Reula for stimulating 
discussions. 
This work was supported by the NSF under grants
PHY-0653369, PHY-0653375, AST-0407070 and AST-0708551 to LSU, 
PHY-0326378 and PHY-0502218
to BYU, and PHY-0325224 and PHY-0643004 to LIU.
Computations were done at BYU, LONI, LSU, and
TeraGrid.

%
%

\bibliographystyle{apsrev}

\end{document}